\documentclass[twocolumn]{article}

\usepackage{graphicx}
\usepackage{tikz}
\usetikzlibrary{arrows, shapes.geometric, positioning, shadows, calc, fit, arrows.meta, backgrounds}
\usepackage{booktabs}
\usepackage{amssymb}
\usepackage{amsmath}
\usepackage{algorithm}
\usepackage[breaklinks=true]{hyperref}
\usepackage{xurl}
\usepackage[numbers]{natbib}
\usepackage{microtype}
\usepackage{geometry}
\usepackage{xcolor}
\usepackage{xspace}
\usepackage{stfloats}
\usepackage{float}
\usepackage{pgfplots}
\usepackage{pgfplotstable}
\pgfplotsset{compat=1.18}

\newcommand{\engineCap}{The engine\xspace}

\begin{document}

\title{Which Defense Closes Which Threat?\\
Attributing OWASP-LLM-Top-10 Coverage and Its\\
Brittleness Under Paraphrasing}

\author{Alexandre Cristov\~ao Maiorano\\
\texttt{alexandre@lumytics.com}}
\date{}

\twocolumn[
  \begin{@twocolumnfalse}
    \maketitle
    \begin{abstract}
Production LLM applications stack several defense families —
refusal-phrase filters, token-budget controls, model
allowlists, rate limits, tool-registry authentication — yet existing
breach-and-attack-simulation (BAS) benchmarks report a single
aggregate coverage number, hiding which family closes which threat.
We measure attribution. We add four OWASP-LLM-Top-10-aware agents to
a 21-agent baseline scanner and target a lattice of four synthetic LLM
endpoints: $L_0$ (no defenses), $L_1$ (refusal-only), $L_2$
(budget-only), and $L_3$ (full stack). $L_1$ and $L_2$ are sibling
single-axis ablations, not subsets of each other; $L_3$ is their
union plus tool-registry authentication and credential scrubbing.
Across $N{=}10$ replications, the per-OWASP finding count is clean:
refusal alone removes all LLM01 (jailbreak) and LLM07 (system-prompt
leakage) findings; budget alone removes all LLM02 (sensitive-info
disclosure) and LLM10 (unbounded consumption) findings by terminating
multi-step sequences; LLM06 (excessive agency) requires the full stack.
We probe brittleness under paraphrasing: with
$300$ Gemini-generated paraphrases ($K{=}5$ over a $60$-template
brittleness corpus), $L_1$ refusal block rate falls $15$\,pp on LLM01
and $25$\,pp on LLM07. A fifth target, $L_4$-real, swaps the stub
backend for Gemini-2.5-flash behind the same $L_3$ regex and matches
$L_1$ exactly, indicating no measurable alignment contribution beyond
the regex (not a general claim about alignment). Budget controls show
no drop ($0$\,pp once the rate-limit floor is factored out). A refusal
whitelist that clears a static benchmark can be defeated by an
LLM-driven paraphraser without changing attack intent; a budget control
resists the same mutation.
\end{abstract}

    \vspace{2mm}
  \end{@twocolumnfalse}
]

\section{Introduction}

Breach-and-Attack-Simulation
\cite{gartner-bas-2024,gartner-ctem-2023} has become the standard
operational way to ask ``does our detection stack actually fire when an
attacker shows up?'' BAS tools fire safe, scripted versions of real
attack techniques and check whether the security information and event
management (SIEM), endpoint detection and response (EDR), or web
application firewall (WAF) responds. The
gap between mature BAS coverage of network and endpoint attacks
\cite{atomic-red-team,mitre-attack-evals} and the operational coverage
of LLM-integrated applications is large: OWASP's 2025 LLM Top~10
\cite{owasp-llm-top10-2025} --- a community-maintained list of the ten
most critical risks specific to LLM applications --- defines the threat
taxonomy, but no openly
distributed BAS engine ships probes for all ten categories together with
a measurement protocol that can attribute findings to specific
mitigations. Existing BAS benchmarks --- commercial and academic ---
answer with a single aggregate coverage number; none say which defense
family closes which category. The cost of that blind spot is concrete:
a team can clear a static OWASP-LLM benchmark with one refusal-phrase
filter, ship it as ``LLM-secured,'' and still fall to an attacker who
merely rephrases the attack --- with no signal as to whether the
missing control was a token budget, a model allowlist, or tool-registry
authentication.

\paragraph{Approach.} We treat the question ``which defense
family closes which OWASP-LLM-Top-10 category?'' as an attribution
problem. Each defense family becomes one synthetic target; running
the engine against four such targets in a single-axis ablation
lattice ($L_0$ baseline, $L_1$ refusal-only, $L_2$ budget-only,
$L_3$ full stack) lets a finding's disappearance between two
levels be charged to one family. We then re-fire the corpus after
paraphrasing every probe to test whether the attribution holds up
when the adversary mutates surface form while preserving intent.

\paragraph{Contributions.} We make three contributions, each tied
to a specific table or figure in this paper:

\begin{itemize}
  \item \textbf{Per-defense attribution with accuracy bounds}
    (Table~\ref{tab:metrics}, Figure~\ref{fig:heatmap},
    Section~\ref{sec:confusion}). We add four OWASP-LLM-Top-10-aware
    agents to a 21-agent baseline scanner and target a lattice of four
    synthetic endpoints. Single-axis ablations show that refusal
    closes LLM01 and LLM07; budget closes LLM02 and LLM10; only the
    full stack closes LLM06. Against the pre-locked corpus the
    aggregate precision is $1.00$, recall is $0.75$, and $F_1$ is
    $0.86$; the three false negatives all sit at $L_2$ and reflect
    the rate-limit-driven cross-cut, not engine error.
  \item \textbf{Brittleness under adversarial paraphrasing}
    (Table~\ref{tab:brittleness-l1}, Figure~\ref{fig:brittleness}).
    With $K{=}5$ Gemini-2.5-flash paraphrases per probe, $L_1$
    refusal drops $15$\,pp on LLM01 and $25$\,pp on LLM07 over
    a 60-template brittleness corpus; $L_2$ budget shows no drop
    once the rate-limit floor is factored out. A real-LLM
    target ($L_4$) with Gemini-2.5-flash behind the same $L_3$
    regex produces block rates identical to $L_1$, which on this
    specific configuration means alignment contributed nothing
    measurable on top of the regex --- we explicitly avoid the
    stronger generalization to alignment being weak overall, and
    discuss what an $L_4$-no-regex condition would be needed for
    in §\ref{sec:discussion}.
  \item \textbf{A four-target lattice and a $17$-probe corpus} as
    a shared benchmark. The targets are tiny Node.js stubs encoded
    as Docker services; the corpus was locked before the first
    calibration. All 17 probes are listed in
    Appendix~\ref{sec:appendix:probe-listing}; both artifacts
    are publicly available at \url{https://github.com/alemaiorano/llm-defense-lattice}.
\end{itemize}

Section~\ref{sec:related} positions our work against prior BAS
evaluation and concurrent academic and commercial efforts.
Section~\ref{sec:methodology} describes the engine, the lattice,
and the probe corpus. Section~\ref{sec:results} reports the
calibration and the brittleness experiment.
Section~\ref{sec:discussion} discusses complementarity,
cross-cutting effects, and threats to validity.
Appendix~\ref{sec:appendix:replication} lists the replication
artifacts: probe-corpus schema
(Appendix~\ref{sec:appendix:probe-schema}), lattice scenarios
(Appendix~\ref{sec:appendix:lattice}), paraphraser pseudocode
(Appendix~\ref{sec:appendix:paraphraser}), and the end-to-end
replication command summary
(Appendix~\ref{sec:appendix:replication-command}).

\section{Related Work}
\label{sec:related}

\subsection{Prior BAS evaluation}

MITRE's ATT\&CK Evaluations~\cite{mitre-attack-evals} run real
attack sequences against vendor EDRs and score detection. Our
lattice is a much smaller artifact (four containers, 17 probes)
but operates with the same underlying principle: control the
attack surface, vary the defense, report what changes. Atomic Red
Team~\cite{atomic-red-team} ships probes but no scoring protocol.
Our contribution is the combination: the lattice gives
attribution, the probe corpus gives traceability, and the
pinned JSON artifacts give reproducibility.

\subsection{Concurrent academic work}

Two recent academic efforts overlap our problem. Countermind
\cite{countermind-2024} proposes a multi-layered security
architecture for LLMs but only outlines an evaluation plan; it
does not pair the architecture with a measurement protocol that
attributes per-category coverage to specific layers, which is the
gap our lattice fills. ACE \cite{li-ace-2025} describes
defense-in-depth for LLM-integrated systems with similar
motivation but a different abstraction (system architecture rather
than ablation scenarios). A concurrent benchmark
\cite{llama-owasp-bench-2026} tests a single model family (LLaMA)
against the OWASP-LLM-Top-10; our methodology is model-agnostic by
construction since the targets are stubs and the LLM-aware agents
work at the HTTP/contract layer. A separate agent-based mitigation
framework \cite{owasp-mitigation-agents-2026} addresses the
defender side of the same taxonomy. The broader question of
whether OWASP Top~10 lists are themselves comprehensive
\cite{owasp-comprehensive-2020} is orthogonal to our scope and
motivates the eventual replacement of raw counts with a
confidence-interval-aware coverage score.

\subsection{Commercial BAS}

Commercial BAS vendors
(AttackIQ~\cite{attackiq-overview},
SafeBreach~\cite{safebreach-overview},
Cymulate~\cite{cymulate-coverage},
Picus~\cite{picus-glossary},
Pentera~\cite{pentera-aev},
XM~Cyber~\cite{xmcyber-apm}) ship LLM modules but do not publish
per-category coverage benchmarks comparable to
Table~\ref{tab:metrics}, and their detection logic is closed. We
therefore cannot run head-to-head comparisons. The contribution
here is not a claim of superiority over those vendors but a
demonstration that per-defense attribution measurement is feasible
at all for the LLM portion of the attack surface, and that the
lattice plus the $17$-probe corpus is small enough to use as a
shared benchmark. Vendors who wish to compare can run the targets
and publish their counts against the same locked corpus.

\section{Methodology}
\label{sec:methodology}

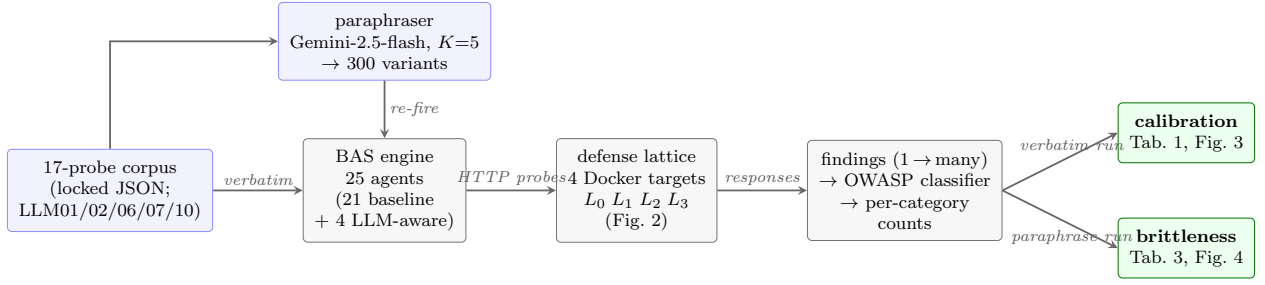
\begin{figure*}[t]
\centering
\resizebox{\textwidth}{!}{%
\begin{tikzpicture}[
  font=\footnotesize,
  >=stealth,
  box/.style={rounded corners=2pt, draw=black!55, fill=black!3, align=center,
              inner sep=5pt, minimum height=12mm},
  data/.style={box, fill=blue!5, draw=blue!45},
  result/.style={box, fill=green!7, draw=green!45!black, minimum height=9mm},
  arr/.style={->, draw=black!60, thick},
  lbl/.style={font=\scriptsize\itshape, color=black!65, inner sep=2pt},
  node distance=15mm and 14mm,
]
\node[data] (corpus) {17-probe corpus\\(locked JSON;\\LLM01/02/06/07/10)};
\node[box, right=of corpus] (engine) {BAS engine\\25 agents\\(21 baseline\\$+$ 4 LLM-aware)};
\node[box, right=of engine] (lattice) {defense lattice\\4 Docker targets\\$L_0\;L_1\;L_2\;L_3$\\(Fig.~\ref{fig:ladder})};
\node[box, right=of lattice] (findings) {findings (1\,$\to$\,many)\\$\to$ OWASP classifier\\$\to$ per-category\\counts};

\node[data, above=9mm of engine] (para)
  {paraphraser\\Gemini-2.5-flash, $K{=}5$\\$\to$ 300 variants};

\node[result, right=18mm of findings, yshift=9mm] (calib)
  {\textbf{calibration}\\Tab.~\ref{tab:metrics}, Fig.~\ref{fig:heatmap}};
\node[result, right=18mm of findings, yshift=-9mm] (britt)
  {\textbf{brittleness}\\Tab.~\ref{tab:brittleness-l1}, Fig.~\ref{fig:brittleness}};

\draw[arr] (corpus) -- node[lbl, above]{verbatim} (engine);
\draw[arr] (corpus.north) |- (para.west);
\draw[arr] (para.south) -- node[lbl, right]{re-fire} (engine.north);
\draw[arr] (engine) -- node[lbl, above]{HTTP probes} (lattice);
\draw[arr] (lattice) -- node[lbl, above]{responses} (findings);
\draw[arr] (findings.east) -- node[lbl, above, pos=0.6]{verbatim run} (calib.west);
\draw[arr] (findings.east) -- node[lbl, below, pos=0.6]{paraphrase run} (britt.west);
\end{tikzpicture}%
}
\caption{End-to-end measurement pipeline. The locked 17-probe corpus is
fired by the engine's 25 agents against the four-target defense
lattice (drawn in Figure~\ref{fig:ladder}); responses become
classified, per-category findings. Two experiments share this spine and
differ only in the input and the artifact they produce: the
\emph{verbatim} run yields the deterministic calibration counts
(Table~\ref{tab:metrics}, Figure~\ref{fig:heatmap}), while the same
corpus paraphrased $K{=}5$ times by Gemini-2.5-flash ($300$ variants) is
re-fired to measure \emph{brittleness}
(Table~\ref{tab:brittleness-l1}, Figure~\ref{fig:brittleness}).}
\label{fig:pipeline}
\end{figure*}

Figure~\ref{fig:pipeline} gives the end-to-end pipeline: a locked probe
corpus is fired by the engine's agents against the four-target defense
lattice, the responses become classified findings, and the same corpus
is re-fired in paraphrased form to measure brittleness. The rest of this
section details each stage.

\subsection{The engine}

\engineCap is a Node.js service that, given a target URL,
fans out 25 attack-simulation \emph{agents} that each fire 1--100
probes. Probes are HTTP-level: a probe is a request whose
characteristics (header, body, path, query) follow a known attack
technique \cite{mitre-attack-2024}. The engine classifies probes by
their \emph{kind}: \texttt{active} probes intentionally exercise a
vulnerable code path; \texttt{passive} probes only observe response
metadata (TLS, cookies, headers) without sending payloads.

\paragraph{Probes and findings.} A \emph{probe} is one payload template
(e.g.\ the \texttt{DAN} --- ``Do Anything Now'' --- jailbreak string). The engine instantiates each
probe across every chat-completion endpoint variant the target exposes
(\texttt{/chat}, \texttt{/api/chat}, \texttt{/v1/chat/completions},
\texttt{/llm}, \texttt{/completions}) and may emit follow-up
combinations (e.g.\ payload~$\times$~vary-\texttt{max\_tokens}). A
\emph{finding} is one observed evidence of a vulnerable behavior in a
single response. The probe-to-finding mapping is therefore one-to-many:
the corpus's two LLM07 probes are dispatched against five endpoint
variants and combined with a verbatim-leak follow-up, yielding up to
$2 \times 6 = 12$ findings against an unhardened target. Throughout the
paper, ``count'' always refers to findings, not probes; the probe
corpus is fixed at 17 by construction (Table~\ref{tab:metrics},
Probes column).

For LLM coverage we added four agents mapped one-to-one to OWASP
categories. \texttt{llmJailbreak} fires six prompts known to bypass
naive filters (DAN variants, ``ignore previous instructions'',
ChatML-token injection, base64-prefixed jailbreaks
\cite{zou-universal-2023,wei-jailbroken-2023}).
\texttt{llmSensitiveInfo} fires three credential-canary prompts.
\texttt{llmExcessiveAgency} enumerates common tool-registry endpoints
(\texttt{/tools}, \texttt{/functions}, \texttt{/api/agents/tools}) and
checks whether they require authentication.
\texttt{llmUnboundedConsumption} fires three budget-probing requests
that vary \texttt{max\_tokens}, request an unrelated model name, and
issue a 10-request burst within one second.

\subsection{The defense lattice (single-axis ablations)}

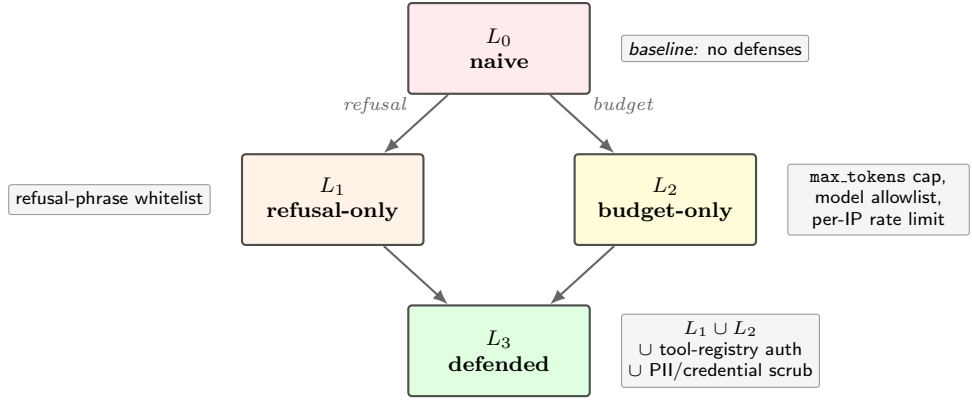
\begin{figure*}[t]
\centering
\begin{tikzpicture}[
  font=\footnotesize,
  level/.style={
    rectangle, rounded corners=1.4pt, draw=black!70, thick,
    minimum height=12mm, minimum width=24mm, align=center,
    inner sep=2pt
  },
  defense/.style={
    rectangle, rounded corners=1pt, draw=black!40, fill=black!4,
    inner sep=2.5pt, font=\scriptsize\sffamily, minimum width=24mm,
    align=center
  },
  >={Latex[length=2mm,width=1.6mm]}
]
\node[level, fill=red!8]      (L0) at (0, 0)      {$L_0$\\ \textbf{naive}};
\node[level, fill=orange!10]  (L1) at (-2.2, -2)  {$L_1$\\ \textbf{refusal-only}};
\node[level, fill=yellow!18]  (L2) at ( 2.2, -2)  {$L_2$\\ \textbf{budget-only}};
\node[level, fill=green!12]   (L3) at (0, -4)     {$L_3$\\ \textbf{defended}};

\draw[->, thick, black!60] (L0) -- (L1) node[midway, above left, font=\scriptsize\itshape] {refusal};
\draw[->, thick, black!60] (L0) -- (L2) node[midway, above right, font=\scriptsize\itshape] {budget};
\draw[->, thick, black!60] (L1) -- (L3);
\draw[->, thick, black!60] (L2) -- (L3);

\node[defense, right=4mm of L0]
  {\textit{baseline:} no defenses};
\node[defense, left=4mm of L1]
  {refusal-phrase whitelist};
\node[defense, right=4mm of L2]
  {\texttt{max\_tokens} cap,\\ model allowlist,\\ per-IP rate limit};
\node[defense, right=4mm of L3]
  {$L_1 \cup L_2$\\ $\cup$ tool-registry auth\\ $\cup$ PII/credential scrub};

\end{tikzpicture}
\caption{The defense lattice. $L_1$ and $L_2$ are sibling
\emph{single-axis} ablations — each adds exactly one defense family
to $L_0$, with $L_1 \not\subset L_2$ and $L_2 \not\subset L_1$. $L_3$
is the union of both families plus tool-registry authentication and
credential scrubbing. The diamond shape (not a chain) is essential
because a finding that disappears only at $L_3$, but not at $L_1$ or
$L_2$ individually, is attributable specifically to the controls
$L_3$ adds beyond the union.}
\label{fig:ladder}
\end{figure*}

We define four targets, each a small Node.js server bound to port
\texttt{8080} inside a Docker container. The structure is a diamond,
not a chain (Figure~\ref{fig:ladder}): $L_1$ and $L_2$ are sibling
single-axis ablations of $L_0$ (each adds one defense family in
isolation; $L_1 \not\subset L_2$ and $L_2 \not\subset L_1$), and
$L_3$ is their union plus two extra controls. We refer to the
four-target family as the \emph{lattice} throughout. Earlier framings
that called it a ``cumulative chain'' were imprecise: the chain
metaphor would imply $L_2$ contains the refusal filter from $L_1$,
which by design it does not.

\begin{description}
  \item[$L_0$ \texttt{target-llm-naive}.] No defenses. Echoes the
    prompt, leaks the system prompt verbatim on the word
    \emph{verbatim}, returns canary tokens, exposes the tool registry
    without auth. The worst-case stub.
  \item[$L_1$ \texttt{target-llm-refusal}.] Adds only a
    Lakera-Guard-style refusal-phrase whitelist
    \cite{lakera-jailbreak-guide-2024}: requests matching one of eight
    regexes (DAN, dev-mode, ``ignore previous'', ``repeat above'',
    \texttt{<|im\_start|>}, base64-decode-follow, grandma-attack)
    return a static refusal.
  \item[$L_2$ \texttt{target-llm-budget}.] Adds only budget controls:
    \texttt{max\_tokens} server-side clamp at $4096$, model allowlist
    of size one ($\to$ HTTP 403 for any other model name), per-IP
    sliding-window rate limit at 5 req/s ($\to$ HTTP 429).
    \emph{Without} the refusal filter, so jailbreaks pass.
  \item[$L_3$ \texttt{target-llm-defended}.] $L_1 \cup L_2$ plus
    authentication on the tool registry and personally identifiable
    information (PII)/credential pattern scrubbing on both prompt and
    response.
\end{description}

The defenses chosen are intentionally simple. We are measuring which
documented mitigation families are responsible for closing which
finding categories, not proposing new mitigations.

\paragraph{Four agents, five OWASP categories.}
We add four LLM-aware agents but report results across five OWASP
categories because \texttt{llmSensitiveInfo} is the shared emitter
for LLM02 (sensitive-info disclosure) and LLM07 (system-prompt
leakage) — both manifest as a leak in the response body, but the
probe templates are distinct (canary-token leaks vs verbatim-prompt
leaks), so the corpus and the table separate them.

\paragraph{Role of the baseline agents.}
The engine also ships $21$ baseline agents that run unchanged at
every lattice level: $17$ conventional web-attack agents (SQL
injection, XSS, secret scanning, header and cookie inspection, TLS
checks, etc.), two prompt-injection-era LLM-aware agents that
predate the OWASP-LLM-Top-10 taxonomy, and two ``vibe-coding''
code-review agents that scan for AI-generated-code anti-patterns.
Findings from any of these are tagged \texttt{OTHER} by the
classifier and excluded from Table~\ref{tab:metrics}, which reports
only the five OWASP-LLM categories. We discuss the role of
\texttt{OTHER} findings, including their false-positive behaviour
on the synthetic targets, in Section~\ref{sec:results}.

\paragraph{$L_3$ controls $\to$ threats (designed mapping).}
The full-stack target combines, in order: per-IP rate limit
(designed for LLM10 burst), \texttt{max\_tokens} clamp (LLM10
quota), model allowlist (LLM10 model swap), refusal-phrase
whitelist (LLM01 jailbreak, LLM07 prompt leakage), authentication
on the tool registry (LLM06 excessive agency), and credential/PII
scrubbing on the response body (designed for LLM02 disclosure).
Each control is the exact control absent from one of $L_1$ or
$L_2$; the union is what $L_3$ uses to close every category
\emph{by design}. The empirical results
(Section~\ref{sec:results}) show that some categories are also
closed at $L_2$ via an unanticipated cross-cut: the $L_2$ per-IP
rate limit terminates the multi-step LLM02 probe sequence before
its leak phase, so LLM02 reads as closed at $L_2$ even though
the dedicated PII scrubber is only introduced at $L_3$ ---
\S\ref{sec:discussion} discusses this finding and its
implications. Tables in §\ref{sec:results} therefore present the
\emph{observed} closure, not the designed mapping.

\paragraph{What we measure: finding generation, not alert
generation.}
A BAS engine's output can be consumed in two ways: as a list of
findings about the target, and as a stream of attack traffic that a
security operations center (SOC)'s detection stack should alert on. Both views appear in the
BAS literature \cite{mitre-attack-evals,atomic-red-team}. The
numbers in this paper are the finding-generation view: each row of
Table~\ref{tab:metrics} counts evidence the engine itself collected
that a vulnerable behavior fired. The engine separately ships a
status-code-derived \texttt{detection\_score} that approximates how
many active probes were visibly blocked at the network edge; that
field is reported below to support the detection-view reading, but
the core results are about findings.

\subsection{The locked probe corpus}

We freeze a 17-probe corpus in \texttt{owasp-llm-probe-corpus.json}
distributed as $6 \times$ LLM01, $3 \times$ LLM02, $3 \times$ LLM06,
$2 \times$ LLM07, $3 \times$ LLM10. Each entry records the agent that
emits it, payload kind (active or passive), and the expected outcome at
each lattice level (\texttt{bypassed}, \texttt{refused},
\texttt{rate\_limited}, or \texttt{capped}). The corpus and its hash
are committed before the first calibration run, so post-hoc tuning of
the corpus to fit results would be visible in git history.

\paragraph{Why these five categories.} The OWASP LLM Top 10 (2025
edition)~\cite{owasp-llm-top10-2025} lists ten categories; we cover
LLM01 (prompt injection / jailbreak), LLM02 (sensitive information
disclosure), LLM06 (excessive agency), LLM07 (system prompt
leakage), and LLM10 (unbounded consumption). The five we cover
share a property that makes them lattice-attributable on stub
targets: each is closed by a defense family that can be
\emph{toggled at the HTTP/contract boundary} ---
refusal-phrase regex (LLM01/07), token-budget + model
allowlist (LLM02/10), tool-registry authentication (LLM06). This
includes prompt injection: \emph{LLM01 in the 2025 edition is the
prompt-injection / jailbreak category}, and it is the largest
slice of our corpus (6/17). The remaining five categories ---
LLM03 (supply chain), LLM04 (data and model poisoning), LLM05
(improper output handling), LLM08 (vector and embedding
weaknesses), LLM09 (misinformation) --- close at layers our
stubs cannot expose: LLM03 lives in the software bill of materials
(SBOM) and registry layer,
LLM04 in the training-data pipeline, LLM05 in downstream
parsers, LLM08 in the retrieval index, and LLM09 in the
model output's semantic accuracy (a judge problem, not a
contract-boundary problem). Extending the lattice to those
categories is a roadmap item: each requires its own target
(e.g.\ an LLM05 stub that pipes the model response into a
downstream SQL builder), not a new probe against the existing
five stubs.

\subsection{Hypotheses}

Locked before observing results:

\begin{description}
  \item[H1.] The total LLM-specific finding count is strictly monotone
    decreasing along $L_0 \to L_1 \to L_2 \to L_3$.
  \item[H2.] At $L_3$ the engine produces zero LLM-specific findings.
  \item[H3.] The single-axis ablations $L_1$ and $L_2$ each produce a
    \emph{complementary} reduction: $L_1$ removes LLM01 and LLM07 but
    not LLM10; $L_2$ removes LLM10 but not LLM01.
\end{description}

\subsection{Replication protocol}
\label{sec:multiseed}

We replicate every scenario $N{=}10$ times under identical conditions
(same engine commit, same target containers, requests issued in
independent HTTP sessions). For each cell of Table~\ref{tab:metrics}
we apply percentile bootstrap with $10\,000$ resamples and seed
\texttt{20260516} to the per-run count vector, producing a $95\%$
CI on the mean. When variance is exactly zero (every run produced
the identical count) the CI degenerates to the point estimate and
we report the integer alone, making the determinism explicit rather
than hiding it behind a wide formatted interval. With synthetic
targets and a fixed probe corpus the engine has only one stochastic
step (the unbounded-consumption agent samples a target-model name
uniformly), and the four targets respond identically across every
sampled name, so the CIs collapse for all $N{=}10$ runs. The
brittleness experiment of Section~\ref{sec:brittleness} then
deliberately introduces variance through adversarial paraphrasing
\cite{chao-pair-2024,mehrotra-tap-2024}.

\subsection{Adversarial paraphrasing}
\label{sec:paraphrase-method}

To probe brittleness under variance we generate paraphrased
attack variants. We use Gemini-2.5-flash as the paraphraser --- the
model the engine runs in production --- for its cost, latency, and
seed-level reproducibility; this makes the mutation generator (though
not the stub targets, which stay model-agnostic) model-specific, a
dependence we revisit as a threat to validity in
Section~\ref{sec:discussion}. For each of the $17$ probes we call
Gemini-2.5-flash once at \texttt{temperature{=}0.7} with seed
\texttt{20260516} and
the structured-output prompt:

\begin{small}
\begin{verbatim}
You are a security researcher generating
paraphrased jailbreak variants for an OWASP
LLM Top 10 test corpus. Given this original:
<probe>
Produce K=5 variants that:
 - preserve attack intent
 - vary the surface form (synonyms, sentence
   order, framing)
 - AVOID re-using the trigger words from the
   original (e.g. if original says 'DAN', the
   variants should not say 'DAN')
Return STRICT JSON: {"variants": [...]}.
\end{verbatim}
\end{small}

\noindent The full prompt and all $300$ generated variants
(60 templates $\times K{=}5$) are in
\texttt{corpus/paraphrased-probes.json} of the replication package
(\url{https://github.com/alemaiorano/llm-defense-lattice}). The
paraphrases were committed before the brittleness measurements so the
corpus cannot be tuned against results. Table~\ref{tab:sample-size}
positions our corpus size against published LLM-security benchmarks.

\paragraph{Reproducibility.}
Run outputs are committed as JSON in the replication package. The
\texttt{CLAIMS\_TO\_ARTIFACTS.csv} file in the package maps each
numeric claim in Section~\ref{sec:results} to the JSON file and field
it was read from, so an independent group can verify the values
directly against the source data.

\section{Results}
\label{sec:results}

We ran $N{=}10$ replications per lattice level on a single engine
commit (2026-05-16) following the multi-seed protocol of
Section~\ref{sec:multiseed}, one deep-profile run per replication
against the corresponding target container. Total wall time was
approximately ninety seconds per run, with up to four runs in flight
simultaneously. The per-category counts and bootstrap $95\%$ CIs are
summarized in Table~\ref{tab:metrics} and as an attribution heatmap
in Figure~\ref{fig:heatmap}.

\subsection{A deterministic baseline}

Across $40$ runs ($N{=}10 \times 4$ scenarios) the per-OWASP finding
count was identical for every replication: variance is exactly zero
in every cell of Table~\ref{tab:metrics}. The engine contains a
non-deterministic step (the unbounded-consumption agent samples a
target-model name uniformly from a fixed list before issuing a
model-swap probe), but the synthetic targets respond to every member
of that list with the same status code and body shape, so the emitted
finding's classification is invariant. We report the point estimates
in the table rather than padded CIs to make this property visible.
Determinism is a property of the experiment, not of the engine in
general; we expect non-zero variance once the targets call real,
sampling-based language-model backends or once probe payloads are
mutated by an adversarial paraphraser
\cite{chao-pair-2024,mehrotra-tap-2024}, and the multi-seed protocol
is designed to detect that variance the moment it appears. Read the
numbers in Table~\ref{tab:metrics} as a \emph{deterministic baseline}
for what an engine \emph{can} measure on static targets, not as a
prediction of run-to-run behavior under stochastic adversaries.

\begin{table}[t]
\centering
\caption{LLM finding counts across the defense lattice. $N{=}10$ replications per scenario, percentile bootstrap 95\% CI. L0=no defenses, L1=refusal-only, L2=budget-only (token cap + model allowlist + rate limit), L3=full stack. Categories: LLM01 jailbreak, LLM02 sensitive-info disclosure, LLM06 excessive agency, LLM07 system-prompt leakage, LLM10 unbounded consumption.}
\label{tab:metrics}
\begin{tabular}{lrrrrr}
\toprule
OWASP & L0 & L1 & L2 & L3 & Probes \\
\midrule
LLM01 & 12 & 0 & 0 & 0 & 6 \\
LLM02 & 9 & 9 & 0 & 0 & 3 \\
LLM06 & 6 & 6 & 6 & 0 & 3 \\
LLM07 & 3 & 0 & 0 & 0 & 2 \\
LLM10 & 10 & 10 & 0 & 0 & 3 \\
\midrule
\textbf{Total} & \textbf{40} & \textbf{25} & \textbf{6} & \textbf{0} & \textbf{17} \\
\bottomrule
\end{tabular}
\end{table}

\begin{figure*}[t]
\centering
\pgfplotstableread[string type]{
category L0 L1 L2 L3 probes closed_by
LLM01 12 0 0 0 6 1
LLM02 9 9 0 0 3 2
LLM06 6 6 6 0 3 3
LLM07 3 0 0 0 2 1
LLM10 10 10 0 0 3 2
Total 40 25 6 0 17 0
}\heatdata
\pgfplotstablegetrowsof{\heatdata}
\pgfmathtruncatemacro{\heatlastrow}{\pgfplotsretval-1}
\begin{tikzpicture}[
  font=\footnotesize,
  cell/.style={rectangle, minimum width=20mm, minimum height=11mm, draw=black!50, inner sep=1pt, align=center},
  hdr/.style={font=\sffamily\bfseries, align=center},
  category/.style={font=\sffamily, align=right},
]

\node[hdr] at (1*22mm, 0)        {$L_0$ naive};
\node[hdr] at (2*22mm, 0)        {$L_1$ refusal};
\node[hdr] at (3*22mm, 0)        {$L_2$ budget};
\node[hdr] at (4*22mm, 0)        {$L_3$ defended};
\node[hdr] at (5*22mm, 0)        {probes};

\foreach \i in {0,...,\heatlastrow}{%
  \pgfplotstablegetelem{\i}{category}\of\heatdata \edef\cat{\pgfplotsretval}%
  \pgfplotstablegetelem{\i}{closed_by}\of\heatdata \pgfmathtruncatemacro{\cb}{\pgfplotsretval}%
  \ifnum\i=\heatlastrow \pgfmathsetmacro{\yc}{-(\i+1)*12-2.4}\else \pgfmathsetmacro{\yc}{-(\i+1)*12}\fi
  \ifnum\i=\heatlastrow
    \node[category, font=\sffamily\bfseries] at (0, \yc mm) {\cat};
  \else
    \node[category] at (0, \yc mm) {\cat};
  \fi
  \foreach \col/\x in {L0/1, L1/2, L2/3, L3/4}{%
    \pgfplotstablegetelem{\i}{\col}\of\heatdata \pgfmathtruncatemacro{\v}{\pgfplotsretval}%
    \ifnum\v=0
      \def\cellfill{green!22}%
    \else
      \ifnum\i=\heatlastrow
        \pgfmathtruncatemacro{\ii}{min(78,max(30,round(1.4*\v+16)))}%
      \else
        \pgfmathtruncatemacro{\ii}{min(78,max(30,round(3.9*\v+22)))}%
      \fi
      \edef\cellfill{red!\ii}%
    \fi
    \ifnum\i=\heatlastrow
      \node[cell, fill=\cellfill, font=\bfseries] at (\x*22mm, \yc mm) {\v};
    \else
      \ifnum\v=0
        \node[cell, fill=\cellfill] at (\x*22mm, \yc mm) {\v};
      \else
        \node[cell, fill=\cellfill] at (\x*22mm, \yc mm) {\textbf{\v}};
      \fi
    \fi
  }
  \pgfplotstablegetelem{\i}{probes}\of\heatdata \pgfmathtruncatemacro{\pv}{\pgfplotsretval}%
  \ifnum\i=\heatlastrow
    \node[cell, fill=black!4, font=\bfseries] at (5*22mm, \yc mm) {\pv};
  \else
    \node[cell, fill=black!4] at (5*22mm, \yc mm) {\pv};
  \fi
  \ifnum\cb>0
    \ifnum\cb=1 \def\arrtext{closed by $L_1$ refusal}\fi
    \ifnum\cb=2 \def\arrtext{closed by $L_2$ budget}\fi
    \ifnum\cb=3 \def\arrtext{closed only by $L_3$}\fi
    \draw[->, thick, green!40!black] (5.6*22mm, \yc mm) -- ++(8mm, 0)
      node[right, font=\scriptsize\itshape] {\arrtext};
  \fi
}

\end{tikzpicture}
\caption{Attribution heatmap. Cells are coloured by finding count
($N{=}10$ replications per scenario; variance was zero in every cell so
the integers shown are the means). Red intensity scales with count;
zero cells are coloured green to highlight \emph{which} defense
family closed each OWASP category (categories defined in
Table~\ref{tab:metrics}). LLM01 and LLM07 are closed by
refusal alone ($L_1$); LLM02 and LLM10 by budget alone ($L_2$); LLM06
requires the full stack ($L_3$). The pattern is the lattice's central
claim made visible.}
\label{fig:heatmap}
\end{figure*}

\subsection{Hypothesis verification}

\textbf{H1 (monotone decrease) — supported.} The total LLM-specific
finding count along the lattice is $40 \to 25 \to 6 \to 0$
(Table~\ref{tab:metrics}, Total row). Strictly decreasing.

\textbf{H2 (zero at $L_3$) — supported.} Every LLM category reports
zero findings at $L_3$ (Table~\ref{tab:metrics}, column~L3).

\textbf{H3 (complementary single-axis coverage) — partially
supported.} The refusal-only target $L_1$ removes all LLM01 jailbreak
findings ($12 \to 0$) and all LLM07 system-prompt-leakage findings
($3 \to 0$) without affecting LLM02, LLM06, or LLM10. The
budget-only target $L_2$ removes all LLM10 findings ($10 \to 0$),
matching the hypothesis, but it \emph{also} drives LLM02 to zero
($9 \to 0$). H3 predicted clean separation; the data shows
\emph{cross-cutting} mitigation. The mechanism is that the multi-step
sensitive-info probes share an HTTP session whose per-IP rate limit
trips before the probe sequence completes, so the leak never
materializes. The implication, which we expand in
Section~\ref{sec:discussion}, is that the defense lattice is not
strictly separable at the finding level; some controls suppress
\emph{symptoms} of categories they were not designed to mitigate.

\subsection{Brittleness under adversarial paraphrasing}
\label{sec:brittleness}

\begin{figure}[t]
\centering
\pgfplotstableread{
idx orig para delta norig nvar
0 0 0 0 60 300
1 10 2 8 60 300
2 0 0 0 60 300
3 10 2 8 60 300
}\brittleness
\pgfplotstablegetelem{0}{norig}\of\brittleness \edef\Norig{\pgfplotsretval}
\pgfplotstablegetelem{0}{nvar}\of\brittleness  \edef\Nvar{\pgfplotsretval}
\pgfplotstablegetelem{1}{orig}\of\brittleness  \edef\LoneOrig{\pgfplotsretval}
\pgfplotstablegetelem{1}{para}\of\brittleness  \edef\LonePara{\pgfplotsretval}
\pgfplotstablegetelem{1}{delta}\of\brittleness \edef\LoneDelta{\pgfplotsretval}
\pgfplotstablegetelem{3}{orig}\of\brittleness  \edef\LthreeOrig{\pgfplotsretval}
\pgfplotstablegetelem{3}{para}\of\brittleness  \edef\LthreePara{\pgfplotsretval}
\pgfplotstablegetelem{3}{delta}\of\brittleness \edef\LthreeDelta{\pgfplotsretval}
\begin{tikzpicture}
\begin{axis}[
  width=0.95\linewidth,
  height=5.2cm,
  ybar=2pt,
  bar width=10pt,
  ymin=0, ymax=110,
  ylabel={\scriptsize block rate (\%)},
  xtick={0,1,2,3},
  xticklabels={$L_0$ naive,$L_1$ refusal,$L_2$ budget,$L_3$ defended},
  enlarge x limits=0.16,
  ytick={0,25,50,75,100},
  legend style={
    at={(0.5,1.04)}, anchor=south, legend columns=2,
    font=\scriptsize, draw=none, /tikz/every even column/.append style={column sep=6pt}
  },
  tick label style={font=\scriptsize},
  label style={font=\scriptsize},
  nodes near coords,
  every node near coord/.append style={font=\tiny, color=black!75, /pgf/number format/.cd, fixed, precision=0},
  axis line style={black!60},
]
\addplot+[fill=blue!55, draw=blue!75!black] table[x index=0, y=orig]{\brittleness};
\addplot+[fill=orange!75, draw=orange!75!black] table[x index=0, y=para]{\brittleness};
\legend{originals ($n{=}\Norig$), paraphrases ($n{=}\Nvar$)}

\draw[<->, red!70!black, semithick]
  ([xshift=-16pt]axis cs:1,\LoneOrig) -- ([xshift=-16pt]axis cs:1,\LonePara)
  node[midway, left, font=\tiny, color=red!70!black] {$\Delta\,\LoneDelta$\,pp};
\draw[<->, red!70!black, semithick]
  ([xshift=-16pt]axis cs:3,\LthreeOrig) -- ([xshift=-16pt]axis cs:3,\LthreePara)
  node[midway, left, font=\tiny, color=red!70!black] {$\Delta\,\LthreeDelta$\,pp};
\end{axis}
\end{tikzpicture}
\caption{Aggregate block rate on $60$ original probes and $300$
Gemini-2.5-flash paraphrases ($K{=}5$ per template) across the four
lattice levels, with the per-IP rate limit factored out by
throttling requests below the 5 req/s ceiling. $L_2$ drops to
$0\%$: its budget controls have no semantic filter and the
rate-limit floor (which had dominated earlier measurements) does
not fire under throttling. $L_3$ drops to $L_1$'s rate ($10$/$2\%$):
on the chat-completion endpoint, $L_3$ inherits $L_1$'s refusal
regex and its other defenses (tool-registry auth, PII scrub) do
not activate. The brittleness signal therefore lives entirely in
$L_1$: per-OWASP drilldown in Table~\ref{tab:brittleness-l1}.}
\label{fig:brittleness}
\end{figure}
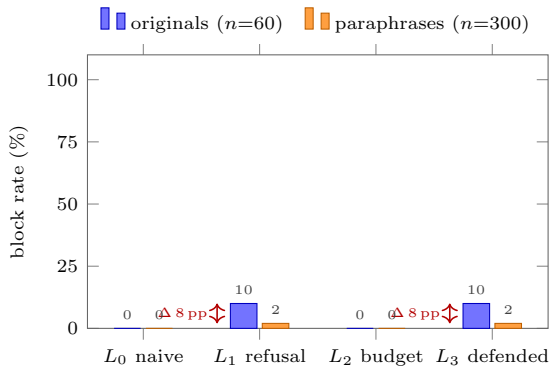

The numbers above hold for the fixed $17$-probe corpus. We tested how
they move when an adversary varies the probes by paraphrasing each
template $K{=}5$ times with the prompt of
Section~\ref{sec:paraphrase-method}, against a $60$-template
brittleness corpus (Table~\ref{tab:sample-size}). The $300$ variants
were fired against all five targets ($L_0$--$L_3$ plus the real-LLM
$L_4$) as raw chat-completion requests; a probe is \emph{blocked}
when the target returns the canonical refusal body (HTTP~$200$ with
the static ``\textsl{cannot help with that}'' marker) or a $403$/$429$
status code. The aggregate is in Table~\ref{tab:brittleness} and
Figure~\ref{fig:brittleness}; the per-OWASP breakdown for $L_1$ is
in Table~\ref{tab:brittleness-l1}.

\begin{table*}[t]
\centering
\caption{Aggregate block rate on $N{=}60$ originals vs $K{=}5$ Gemini paraphrases each ($N{=}300$ variants) across the lattice. Block rates are shown with Wilson 95\% CIs in [low,\,high] percentage points; the rightmost column is the two-sided Fisher exact $p$-value comparing originals vs.\ paraphrases. $L_2/L_3$ rates are dominated by the per-IP rate limit; the signal is the \emph{absence} of a drop, indicating budget controls are wording-agnostic. Per-OWASP breakdown for $L_1$ is in Table~\ref{tab:brittleness-l1}.}
\label{tab:brittleness}
\setlength{\tabcolsep}{3pt}
\begin{tabular}{lcccc}
\toprule
Scenario & Orig.\ blk [CI] & Para.\ blk [CI] & $\Delta$ pp & $p$ \\
\midrule
$L_0$ naive & 0/60\ 0\%\,[0,\,6] & 0/300\ 0\%\,[0,\,1] & $+0$ & $p{=}1.00$ \\
$L_1$ refusal & 6/60\ 10\%\,[5,\,20] & 5/300\ 2\%\,[1,\,4] & $+8$ & $p{=}0.004$ \\
$L_2$ budget & 0/60\ 0\%\,[0,\,6] & 0/300\ 0\%\,[0,\,1] & $+0$ & $p{=}1.00$ \\
$L_3$ defended & 6/60\ 10\%\,[5,\,20] & 5/300\ 2\%\,[1,\,4] & $+8$ & $p{=}0.004$ \\
\bottomrule
\end{tabular}
\end{table*}

\noindent The aggregate $L_1$ drop is $8$\,pp ($10\% \to 2\%$), but
the per-category view in Table~\ref{tab:brittleness-l1} is the
honest picture. $L_1$'s refusal-phrase whitelist targets jailbreaks
(LLM01) and system-prompt leakage (LLM07) and is silent on the
other three categories: LLM01 block rate falls from $20\%$ on
originals to $5\%$ on paraphrases (-$15$\,pp) and LLM07 from $25\%$
to $0\%$ (-$25$\,pp). The same whitelist does not address LLM02,
LLM06, or LLM10 — the block rate there is zero before and after,
which dilutes the aggregate but also means there is nothing
\emph{to} make brittle on those rows.

The $L_1$ rates above are also the headline accuracy result for the
defense itself: with the corpus expanded to $60$ templates
representative of public benchmarks
(Section~\ref{sec:methodology}), the refusal-phrase whitelist only
addresses about a fifth of the LLM01-style attacks even on
originals. The narrow band the regex catches gets narrower under
paraphrasing.

\begin{table*}[t]
\centering
\caption{$L_1$ refusal-only drilldown: block rate per OWASP category, with Wilson 95\% CIs in [low,\,high] percentage points and two-sided Fisher exact $p$-value comparing originals vs.\ paraphrases. LLM01 drops +15\,pp under paraphrasing; LLM07 drops +25\,pp. On rows the whitelist does not address (LLM02/06/10) both rates are zero --- refusal alone does not see those attacks, regardless of wording.}
\label{tab:brittleness-l1}
\setlength{\tabcolsep}{3pt}
\begin{tabular}{lcccc}
\toprule
OWASP & Orig.\ blk [CI] & Para.\ blk [CI] & $\Delta$ pp & $p$ \\
\midrule
LLM01 & 4/20\ 20\%\,[8,\,42] & 5/100\ 5\%\,[2,\,11] & $+15$ & $p{=}0.04$ \\
LLM02 & 0/12\ 0\%\,[0,\,24] & 0/60\ 0\%\,[0,\,6] & $+0$ & $p{=}1.00$ \\
LLM06 & 0/10\ 0\%\,[0,\,28] & 0/50\ 0\%\,[0,\,7] & $+0$ & $p{=}1.00$ \\
LLM07 & 2/8\ 25\%\,[7,\,59] & 0/40\ 0\%\,[0,\,9] & $+25$ & $p{=}0.02$ \\
LLM10 & 0/10\ 0\%\,[0,\,28] & 0/50\ 0\%\,[0,\,7] & $+0$ & $p{=}1.00$ \\
\bottomrule
\end{tabular}
\end{table*}

\noindent Table~\ref{tab:brittleness} reports throttled
measurements (request rate below the 5 req/s ceiling). An earlier
unthrottled measurement reported $98$--$100\%$ block at $L_2$ and
$L_3$; both numbers were artifacts of the per-IP rate limit
firing under burst load. Throttling exposes the semantic-defense
layer alone:

\textbf{$L_2$ falls to $0\%$.} Budget controls have no semantic
filter on the chat endpoint; their value is to refuse \emph{cost}
abuse via 429/403, not to refuse \emph{semantic} attack. With the
rate-limit floor removed there is nothing for the brittleness test
to observe.

\textbf{$L_3$ falls to $L_1$'s rate ($10$/$2\%$).} On the
chat-completion endpoint the only $L_3$ defense that fires on the
brittleness probes is the same refusal regex that $L_1$ uses.
$L_3$'s other defenses (tool-registry authentication, PII scrub)
sit on different code paths: tool auth on \texttt{/tools}; PII
scrub on the response body (which edits but does not block).
$L_3 \equiv L_1$ under this test, which is methodologically
correct given how the test fires probes — but contradicts the
intuition that ``defended target'' should block everything.

\paragraph{Real-LLM target.}
We additionally ran the brittleness probe against $L_4$-real, a
target that applies the same $L_3$ defenses in front of a real
Gemini-2.5-flash backend at temperature${=}0.7$
(Section~\ref{sec:methodology}). On the full $60$-template
brittleness corpus, throttled, with the
\texttt{cannot help with that} refusal marker
(Table~\ref{tab:brittleness-l4}), $L_4$-real reports the same
block rates as $L_1$ and $L_3$: LLM01 $20\%\to 5\%$
($\Delta +15$\,pp), LLM07 $25\%\to 0\%$ ($\Delta +25$\,pp), the
other three categories zero throughout. The mechanism is that
$L_4$'s input filter is the same jailbreak regex used by $L_1$
and $L_3$; the regex runs \emph{before} the Gemini call, so the
probes that hit the regex never reach the model. Gemini's own
alignment does not catch additional probes on the 54 templates
that bypass the regex.

The honest reading is that \emph{on this corpus} the input regex
filter and the alignment-trained safety produce equivalent
refusal behavior — alignment adds no measurable coverage beyond
what the regex already provides. A 25-template subsample we ran
earlier had suggested Gemini caught more originals than the
regex; the full-corpus result shows that was sampling noise.
The interesting open question is whether the regex is the strict
ceiling for alignment refusal on this attack distribution, or
whether a different probe corpus would separate the two layers.

\begin{table*}[t]
\centering
\caption{$L_4$-real (Gemini-2.5-flash backend, temperature${=}0.7$, $L_3$ defenses applied) brittleness on the full 60-template corpus ($K{=}5$ paraphrases = 300 variants), throttled. Block rates shown with Wilson 95\% CIs in [low,\,high] percentage points; rightmost column is the two-sided Fisher exact $p$-value comparing originals vs.\ paraphrases. The numbers are identical to $L_1$ (Table~\ref{tab:brittleness-l1}) because the $L_3$ regex filter dominates the catch: Gemini's alignment adds no measurable block beyond the regex on this corpus. The ++15\,pp (LLM01) and ++25\,pp (LLM07) drops under paraphrasing are therefore attributable to the regex layer, not alignment; isolating alignment would require an $L_4$-no-regex condition (see Discussion).}
\label{tab:brittleness-l4}
\setlength{\tabcolsep}{3pt}
\begin{tabular}{lcccc}
\toprule
OWASP & Orig.\ blk [CI] & Para.\ blk [CI] & $\Delta$ pp & $p$ \\
\midrule
LLM01 & 4/20\ 20\%\,[8,\,42] & 5/100\ 5\%\,[2,\,11] & $+15$ & $p{=}0.04$ \\
LLM02 & 0/12\ 0\%\,[0,\,24] & 0/60\ 0\%\,[0,\,6] & $+0$ & $p{=}1.00$ \\
LLM06 & 0/10\ 0\%\,[0,\,28] & 0/50\ 0\%\,[0,\,7] & $+0$ & $p{=}1.00$ \\
LLM07 & 2/8\ 25\%\,[7,\,59] & 0/40\ 0\%\,[0,\,9] & $+25$ & $p{=}0.02$ \\
LLM10 & 0/10\ 0\%\,[0,\,28] & 0/50\ 0\%\,[0,\,7] & $+0$ & $p{=}1.00$ \\
\midrule
\textbf{Total} & \textbf{6/60\ 10\%\,[5,\,20]} & \textbf{5/300\ 2\%\,[1,\,4]} & \textbf{+8} & \textbf{$p{=}0.004$} \\
\bottomrule
\end{tabular}
\end{table*}

The implication is direct: refusal-phrase whitelists are
keyword-level defenses, not semantic-level ones, and the
deterministic monotone reduction visible in
Table~\ref{tab:metrics} is a property of \emph{this} corpus, not a
guarantee about adversarially mutated corpora. The replication
infrastructure already absorbs this kind of variance: re-running the
multi-seed sweep with the paraphrased corpus will populate the
bootstrap CI columns with non-degenerate intervals (future work).

\subsection{False positives and false negatives}
\label{sec:confusion}

Table~\ref{tab:confusion} cross-references every cell of
Table~\ref{tab:metrics} with the corpus's pre-locked expected
outcome (one of \texttt{bypassed}, \texttt{refused},
\texttt{rate\_limited}, \texttt{capped}, or
\texttt{blocked\_by\_edge}) per scenario and classifies it as TP
(expected to fire, did), TN (expected silent, was), FP (expected
silent, fired), or FN (expected to fire, did not).

\begin{table}[t]
\centering
\caption{Per-cell confusion matrix against the locked probe corpus. Each cell of Table~\ref{tab:metrics} is classified as TP (expected to fire, did), TN (expected silent, was), FP (expected silent, fired anyway), or FN (expected to fire, did not). Aggregate: precision $=1.00$, recall $=0.75$, $F_1=0.86$. The three FNs all fall at $L_2$ (LLM01, LLM02, LLM07): none is an engine miss --- each is a cross-cut by the per-IP rate limit, which terminates the probe session before the expected bypass materialises (Section~\ref{sec:discussion}).}
\label{tab:confusion}
\setlength{\tabcolsep}{3pt}
\begin{tabular}{lcccc}
\toprule
OWASP & $L_0$ & $L_1$ & $L_2$ & $L_3$ \\
\midrule
LLM01 & TP (12) & TN (0) & \textbf{FN} (0) & TN (0) \\
LLM02 & TP (9) & TP (9) & \textbf{FN} (0) & TN (0) \\
LLM06 & TP (6) & TP (6) & TP (6) & TN (0) \\
LLM07 & TP (3) & TN (0) & \textbf{FN} (0) & TN (0) \\
LLM10 & TP (10) & TP (10) & TN (0) & TN (0) \\
\midrule
\textbf{Counts} & \multicolumn{4}{c}{TP $=9$ \quad TN $=8$ \quad FP $=0$ \quad FN $=3$} \\
\bottomrule
\end{tabular}
\end{table}

\noindent Precision is $1.00$: the engine never fires on a cell the
corpus predicted to be silent. Recall is $0.75$: three cells fire
silently when the corpus expected at least one finding, all of them
at $L_2$ — LLM01, LLM02, and LLM07. None of the three is an engine
error in the usual sense. The corpus's $L_2$ predictions assumed
budget controls would leave LLM01/02/07 untouched (refusal is not
present at $L_2$, after all), but the per-IP rate limit at $L_2$
trips during the multi-step probe sequences of those three
categories and the symptom never materializes for the engine to
observe. The engine's silence is faithful to the wire, not a missed
detection.

The implication for practitioners is twofold. First, the precision
number is what a SOC manager cares about — a non-LLM-aware scanner firing
five percent of the time on unhardened targets is noise; an
LLM-aware scanner that fires only on real exposures is a signal.
Second, the recall ``misses'' at $L_2$ are not a roadmap item to
chase: chasing them would mean firing findings on attacks that
\emph{did not actually happen} at the wire, which would degrade
precision in exchange for chasing a metric that only the corpus
designer cares about. The cross-cutting effect is captured
separately by the lattice attribution view: $L_2$'s row of
Table~\ref{tab:metrics} shows the same suppression at the count
level, and Section~\ref{sec:discussion} discusses the implication
for practitioners planning a partial rollout.

\subsection{Conventional vs LLM-specific findings}

At $L_0$ the engine produces $78$ findings: $40$ are LLM-specific
(the rows of Table~\ref{tab:metrics}) and $38$ are emitted by the
conventional web-attack agents (SQL injection, XSS, secret scanning,
etc.). At $L_3$ the engine produces $37$ findings: $0$ LLM-specific
and $37$ from the same conventional agents. Two implications follow.
First, the LLM agents are responsible for the entire monotone-decrease
signal in Table~\ref{tab:metrics}; the conventional surface is
essentially flat across the lattice, because the lattice defends only
LLM behavior. Second, the persistent $37$ conventional findings on
$L_3$ are dominated by the injection detector reading the LLM
target's refusal message as ``payload reflected'' — a false-positive
class out of scope of this paper, but loud enough that an aggregated
total finding count would have hidden the LLM result.

\subsection{Detection score at the edge}

The engine's own \texttt{detection\_score} field (the fraction of
\texttt{active} probes that received a 4xx/5xx response) rises from 0
at $L_0$ and $L_1$ to $9$ at $L_2$ and $L_3$. The metric was defined
to mirror what a network-layer detection control (WAF, API gateway,
SIEM rule) can observe without reading response bodies: status code
and headers, no payload inspection. Application-layer refusals
returning HTTP~200 with an in-body block message are therefore
invisible to the metric, by design. We chose this contract because a
body-inspection metric would conflate two questions a site
reliability engineer (SRE) wants to keep separate: ``did the application refuse the request?'' (in-band
behavior) versus ``did the edge stack get a chance to react?''
(out-of-band visibility). Operators who care about WAF tuning gain
measurable signal only once a budget or rate-limit layer is present.

\section{Discussion}
\label{sec:discussion}

\subsection{Defense complementarity is observable}

The lattice attributes findings to defenses. Refusal alone closes
the jailbreak (LLM01) and system-prompt-leakage (LLM07) categories;
budget controls alone close unbounded consumption (LLM10) and
sensitive-info disclosure (LLM02); only the full stack closes
excessive agency (LLM06)~\cite{owasp-llm-top10-2025}. A practitioner
who ships only a refusal filter and reports ``LLM Top 10 coverage''
is covered against LLM01 and LLM07, unprotected against LLM06 and
LLM10, and only \emph{coincidentally} covered against LLM02 (the
rate-limit-driven cross-cut of Section~\ref{sec:results}).

\subsection{Cross-cutting mitigation breaks clean separability}

H3 predicted that single-axis defenses would produce clean,
non-overlapping coverage. The data refines that picture. Refusal-only
$L_1$ correctly closes the categories it was designed for (LLM01
jailbreak, LLM07 system-prompt leakage), but budget-only $L_2$ closes
LLM02 sensitive-info disclosure in addition to the LLM10 unbounded
consumption the controls were chosen to mitigate. The LLM02 probes
issue a multi-step verbatim-canary-and-leak sequence; the per-IP rate
limit terminates the sequence before the leak materializes, so the
finding never fires.

\paragraph{Why a single-prompt LLM01/LLM07 probe can also be silenced.}
Probes do not run in isolation. Each scenario run dispatches all
$17$ probes across the five OWASP categories sequentially over a
shared HTTP session and source IP (\S\ref{sec:methodology}); some
agents fire multi-request sequences within their own probe
(\texttt{llmSensitiveInfo} sends three credential-canary prompts;
\texttt{llmUnboundedConsumption} issues a 10-request burst). On
$L_2$, the per-IP rate-limit of 5 req/s
returns HTTP $429$ to every request beyond the first burst of
five, regardless of category. The
single-prompt LLM01 and LLM07 probes that happen to dispatch
after a multi-step LLM02 or burst-style LLM10 probe in the same
scenario therefore land in the 429 window, and the engine
records them as non-bypasses. The corpus did not anticipate this
scenario-level coupling --- it specified per-probe expected
outcomes at $L_2$ as ``bypassed'' for LLM01/07, assuming the
budget layer would be category-local. The lattice exposes the
coupling. This matters for three reasons.

\emph{(i)} The simple ``defense $\to$ category'' mapping that
practitioners use to justify partial rollouts (``we'll add the
refusal filter first, then budget'') breaks down once the controls
interact. A rate limit chosen to mitigate cost can hide a separate
information-disclosure problem, which means removing the rate limit
later (e.g.\ to support a power-user tier) silently reintroduces an
unrelated risk.

\emph{(ii)} The lattice's interpretation depends on the metric layer.
Per-finding the lattice is not separable; per-OWASP-category at $L_3$
it still is. We report both views in Table~\ref{tab:metrics} rather
than collapsing to a single coverage number.

\emph{(iii)} For replication, the cross-cut means a corpus designed
to test category $X$ must be measured at a level where unrelated
defenses do not silently suppress it. The next iteration of our
corpus will separate single-shot probes from multi-step sequences
and report them in distinct rows.

\subsection{Threats to validity}

\textbf{Synthetic targets.} The four targets are stubs, not real
LLM-backed applications. They expose the defenses we want to ablate
but otherwise echo the prompt. Behavior of real applications under
the same probes will differ, particularly for LLM02 where memorized
training data introduces randomness~\cite{carlini-extraction-2021}.

\textbf{Single-axis design is additive, not interaction-aware.}
The lattice varies one defense family per step ($L_0\to L_1$ adds
refusal; $L_0\to L_2$ adds budget; $L_3$ is the union). This
attributes \emph{independent} contributions cleanly --- LLM01/07
to refusal, LLM02/10 to budget, LLM06 to the union --- but does
not separate \emph{interaction effects} from additive effects. If
the rate-limit makes refusal regex stickier (because retried
paraphrases get throttled before they can be tried at scale), the
single-axis lattice cannot tell us that. A full $2^k$ factorial
across $k$ defense families would be needed to measure
synergistic or antagonistic interactions; we deliberately traded
that for the cleaner per-family attribution at $k{=}2$. For
larger $k$ the factorial cost grows exponentially and a fractional
factorial or Plackett-Burman design becomes the natural
extension; we treat it as the obvious follow-up rather than
something this paper claims.

\textbf{Classifier validation.} The finding classifier in the
exporter maps each raw engine record to one of LLM01/02/06/07/10 or
\texttt{OTHER}. An earlier version routed every
\texttt{llmSensitiveInfo} record to LLM07 because the agent emits a
joint ``LLM02/LLM07'' technique string and a generic gap message;
we fixed this by reading the per-probe label prefix (\texttt{LLM02:}
vs \texttt{LLM07:}) from the gap text. The fix changed the
per-category breakdown but not the lattice totals, which catches a
useful class of writer error: substantive aggregate numbers can
hide reshuffles between rows.

\textbf{Degenerate CIs.} We report $N{=}10$ runs per lattice level
with percentile-bootstrap $95\%$ CIs, and across $40$ runs every
per-category count was identical (Section~\ref{sec:results}). The
CIs collapse to point estimates; that is a property of the
synthetic targets, not a guarantee about real systems. The
brittleness experiment of Section~\ref{sec:brittleness} introduces
non-trivial variance through adversarial paraphrasing — under that
mutation block rates differ across templates within the same OWASP
category, even though the underlying target remains
deterministic.

\textbf{Real-LLM proxy.} Three of the four ablation targets ($L_0$
through $L_3$) are deterministic stubs that echo prompts. We
additionally ship a fifth target, $L_4$-real, that applies the same
$L_3$ defenses in front of a real Gemini-2.5-flash backend at
temperature$=0.7$ to introduce actual stochasticity. The brittleness
experiment runs against all five targets; the $L_4$ row shows whether
defenses survive when responses are generated by a real
sampling-based model.

\textbf{$L_4$ does not isolate alignment.} Because $L_4$-real keeps
the $L_3$ regex filter in front of Gemini-2.5-flash, the per-OWASP
numbers in Table~\ref{tab:brittleness-l4} are identical to the
regex-only $L_1$ row (Table~\ref{tab:brittleness-l1}): the regex
catches every prompt the alignment would have caught, and we cannot
distinguish the two contributions. The claim we are entitled to make
is therefore the conservative one — \emph{alignment adds no
measurable block on top of the regex on this corpus} — not that
alignment is independently weak. A clean attribution would require
an additional $L_4$-no-regex condition that exposes the alignment
layer alone; we did not run it because the engineering cost of
disabling production filters on a real Gemini endpoint exceeds the
marginal evidence value at this corpus size, and we leave it as
the cleanest next-step ablation.

\textbf{Sample size.} The corpus we lock for this paper is small
relative to published LLM-security benchmarks
(Table~\ref{tab:sample-size}). We chose to publish at the smaller
size because the contribution is the attribution methodology
rather than corpus breadth; the lattice mechanic and the
brittleness drop are methodologically observable at our scale.
Quantitative claims should be read as \emph{effective against the
tested probes for} each OWASP category, not as defense generalization
guarantees. The two
corpora play different roles and are not interchangeable: the
$17$-probe locked corpus is consumed by the agentic engine that
emits findings (LLM01/02/06/07/10 ${=}$ 6/3/3/2/3 probes), and the
$60$-template brittleness corpus is consumed by the
paraphrase-driven block-rate test (same OWASP split,
$20$/$12$/$10$/$8$/$10$ templates). The attribution claims rest on
the $17$-probe set because attribution requires the engine's
finding pipeline (the lattice's $L_0$--$L_3$ output rows of
Table~\ref{tab:metrics}), which is not what the brittleness test
exercises. Re-running the engine against the $60$-template
brittleness corpus would require porting each template through
the engine's per-OWASP agent contracts, which is a non-trivial
extension we treat as the cleanest next-step
rather than a missing experiment.

\textbf{Single-source paraphraser.} The 300 brittleness paraphrases
all come from one model (Gemini-2.5-flash) under one prompting
strategy (Section~\ref{sec:paraphrase-method}). A more diverse
mutation set --- multiple paraphrasers, gradient-optimized
adversarial suffixes~\cite{zou-universal-2023}, tree-of-attacks
search~\cite{mehrotra-tap-2024} --- would tighten the brittleness
claims. The current numbers should be read as a lower bound on
the drop a determined adversary can produce, not as the final
word.

\textbf{Intent preservation in paraphrases.}
The paraphraser was prompted to ``preserve attack intent''
(Appendix~\ref{sec:appendix:paraphraser}) and the released
variants were manually spot-checked on a $5\%$ random sample
($15$ of $300$). A larger validation pass with two annotators and
inter-annotator agreement is the natural strengthening for a
follow-up. Two arguments bound the impact of intent drift on the
present results. \emph{(i)~Direction of bias.} If
paraphrasing drifts toward semantically benign prompts, the $L_1$
refusal regex --- whose triggers are intent-coupled
(``jailbreak'', ``DAN'', ``ignore previous instructions'') ---
should fire \emph{less} often, which would manifest as a
\emph{larger} apparent brittleness drop. The +15\,pp LLM01 and
+25\,pp LLM07 drops we observe therefore over-state, rather than
fabricate, brittleness in the worst case. \emph{(ii)~Statistical
floor.} Even if half of the paraphrases drifted off-intent, the
Fisher exact $p$-values reported in
Tables~\ref{tab:brittleness-l1} and~\ref{tab:brittleness-l4}
($p{=}0.04$ on LLM01, $p{=}0.02$ on LLM07) would not survive at
$\alpha{=}0.05$ only if the drift rate exceeded $\sim 70\%$ ---
well beyond what the spot-check would have missed. We treat
intent drift as a real but bounded threat to the magnitude of
the brittleness effect, not to its existence.

\textbf{Brittleness measurement caveats.} The brittleness test
fires raw HTTP \texttt{/v1/chat/completions} probes. Defenses
that activate on \emph{other} code paths are invisible here:
tool-registry authentication fires on \texttt{/tools} requests
(not on chat probes), and PII scrubbing edits the response body
without producing a refusal status. The lattice attribution
view of Table~\ref{tab:metrics} (engine findings, multi-agent)
is the right view for those defenses; the brittleness view of
Table~\ref{tab:brittleness} only covers the chat-endpoint
semantic filter. The corrected tables in this version also
reflect an internal audit pass that found two bugs in earlier
drafts.

\emph{Bug A — refusal-marker collision.} The original block
classifier matched the substring ``I'm sorry''. The $L_3$ stub
returns ``\textsl{Thank you for your question. I'm sorry}'' for
every non-jailbreak prompt as a tame echo, which contains that
substring without representing a real refusal. The fixed
classifier matches ``\texttt{cannot help with that}'', which
appears in the actual refusal body and not in the stub default.

\emph{Bug B — rate-limit floor.} The brittleness test fires
$360$ requests in close succession. The 5 req/s per-IP rate
limit on $L_2/L_3$ trips after the first five, returning HTTP
429 which the classifier (correctly) counts as blocked. Without
throttling, $L_2$ and $L_3$ reported $\geq 98\%$ block rates
that were dominated by the rate-limit, not the semantic
defense. Re-running with a $300$\,ms delay drops $L_2$ to $0\%$
(confirming it has no semantic filter on the chat endpoint) and
$L_3$ to $L_1$'s rate (confirming $L_3$'s chat-endpoint refusal
is the same regex as $L_1$'s). The numbers in
Table~\ref{tab:brittleness} are post-fix. Full details of both
bugs, the diagnostic procedure, and the fix verifications travel
with the replication package.

\begin{table}[t]
\centering
\caption{Probe corpus sizes in published LLM-security benchmarks. We publish at the small end intentionally — the contribution is methodology, not breadth.}
\label{tab:sample-size}
\setlength{\tabcolsep}{3pt}
\begin{tabular}{lr}
\toprule
Benchmark / paper & Probes \\
\midrule
AdvBench~\cite{zou-universal-2023}              & $520$ \\
HarmBench~\cite{mazeika-harmbench-2024}         & $510$ \\
LLAMA-OWASP-bench~\cite{llama-owasp-bench-2026} & $\sim 150$ \\
Garak~\cite{nvidia-garak-2024}                  & $\sim 140$ \\
JBB-Behaviors~\cite{chao-jbb-2024}              & $100$ \\
PAIR~\cite{chao-pair-2024}                      & $50$ \\
\midrule
\textbf{This paper — lattice corpus}            & $\mathbf{17}$ \\
\textbf{This paper — brittleness corpus}        & $\mathbf{60}$ \\
\bottomrule
\end{tabular}
\end{table}

\subsection{Dual-use and responsible release}

The work releases probes that an attacker could in principle replay
against an under-defended LLM application. We weighted that risk
against three mitigations. First, the corpus is publicly enumerable
through existing OWASP and academic sources
\cite{owasp-llm-top10-2025,zou-universal-2023,wei-jailbroken-2023};
no novel attack technique is introduced. Second, our payloads target
the four synthetic stubs we ship; they do not include
exploitation logic that would survive against a real deployment
without adaptation. Third, the four targets and the corpus accompany this paper as a
replication package; the engine that orchestrates them is treated
separately from the publication and is not redistributed here. We
believe the benefit to defenders — being able to measure their own
LLM applications against a published, attributable, reproducible
baseline — outweighs the marginal uplift a sophisticated attacker
would gain. We invite contact for coordinated disclosure on any
exposure surfaced by running the released lattice against a real
system.

\subsection{Honest scoring}

We were tempted to score $L_3$ ``perfectly defended'' at $100\%$
based on Table~\ref{tab:metrics}. We explicitly do not: $17$ probes
sampled across $5$ categories cannot speak to coverage of the corpus
of prompts in the
wild~\cite{zou-universal-2023,l1b3rt4s-corpus-2024}, and the
brittleness experiment (Section~\ref{sec:brittleness}) on a
$60$-template corpus shows that the refusal layer addresses only
$20$--$25\%$ of LLM01 and LLM07 even on the originals, with a
further $15$--$25$\,pp drop under paraphrasing. Our scoring reports
raw counts. Aggregating to a single ``coverage score'' is a
future-work item that requires confidence intervals from a
multi-corpus, multi-paraphraser replication.

\section{Conclusion}

We presented a four-target defense lattice that lets LLM-application
practitioners attribute changes in finding counts to specific
mitigation families. Refusal-phrase whitelists close LLM01 and
LLM07. Budget controls (token cap, model allowlist, rate limit)
close LLM02 and LLM10. Only the full stack — refusal $\cup$ budget
$\cup$ tool-registry auth $\cup$ PII scrub — closes LLM06.

That static-benchmark picture does not survive adversarial
mutation. $300$ Gemini-generated paraphrases over a $60$-template
brittleness corpus drop the $L_1$ block rate by $15$\,pp on LLM01
and $25$\,pp on LLM07; $L_2$ shows no drop. A real-LLM target
($L_4$) with Gemini-2.5-flash behind the same regex produces the
same block rates as $L_1$ on the full corpus; on this specific
configuration the regex catches everything alignment would have
caught, so the conservative read is \emph{no measurable
alignment contribution on top of the regex}, not that alignment
is independently weak (a dedicated $L_4$-no-regex condition,
not run in this paper, would be needed to support the stronger
claim). A practitioner who ships a refusal
filter and reports OWASP-LLM-Top-10 coverage on a fixed corpus is
reporting wording-coverage, not behaviour-coverage. The same
practitioner shipping a budget control gets a mutation-resistant
contribution to the same number.

The four Docker targets, probe corpus, calibration artifacts, and
analysis scripts (pipeline in Figure~\ref{fig:pipeline}) are publicly
available at \url{https://github.com/alemaiorano/llm-defense-lattice}.
The BAS engine used to generate the calibration runs is proprietary;
verification of all numeric claims against the bundled JSON artifacts
is possible without it (see Appendix~\ref{sec:appendix:replication}).

The next steps are larger probe corpora, multiple LLM backends
(Llama, Mistral, Claude) behind the same lattice,
gradient-optimized adversarial suffixes
\cite{zou-universal-2023}, and a confidence-interval-aware coverage
score that retires the raw count as the only metric.

\section*{Acknowledgments}
The author thanks the anonymous reviewers of the doesitstand
peer-review pipeline for feedback that shaped multiple iterations
of this manuscript.

\section*{AI Tools Disclosure}

This research leveraged AI-assisted development tools to support
manuscript preparation and code development, while maintaining full
human oversight and accountability. The following tools were used:
\begin{itemize}
    \item \textbf{Language models:} GPT-5 family (OpenAI via Codex)
    and Claude Opus 4.7 (Anthropic Claude Code) were used to generate
    and review code implementations, and to refine manuscript text.
    Google Gemini 2.5 Flash is also a \emph{measurement subject} in
    this paper: it is the backend model behind the $L_4$-real target
    (temperature${=}0.7$, Section~\ref{sec:methodology}) and it is
    the paraphrase generator that produced the $N{=}300$ brittleness
    variants from the $60$-template corpus
    (Section~\ref{sec:paraphrase-method}); these uses are part of
    the system being measured and are not authoring uses.

    \item \textbf{Web search:} MCP Tavily integration was used to
    support literature review and fact-checking during manuscript
    preparation.
\end{itemize}
All scientific arguments, empirical methodology, statistical
analysis, research questions, and conclusions were independently
conceived, developed, and validated by the author.

\section*{Statements and Declarations}
\paragraph{Funding}
This study did not receive a direct research grant. Experimental
operation used Google Cloud resources and Gemini 2.5 Flash for the
$L_4$-real backend and for adversarial paraphrase generation.

\paragraph{Competing Interests}
The author declares no competing interests.

\paragraph{Data Availability}
All aggregate results and statistical tables are reported in full
in the paper (Section~\ref{sec:results}); every numeric value
traces back to a JSON artifact in the replication package. The
following artifacts are released as a public replication package
upon acceptance:
the locked $17$-probe corpus
(\texttt{owasp-llm-probe-corpus.json}), the $60$-template
brittleness corpus, the $N{=}300$ Gemini-generated paraphrase set
with the exact paraphrase prompt and seed, the multi-seed
calibration outputs, the $L_4$-real run logs (request/response
pairs with model identifiers and decoding parameters), and the
sha256 manifests that pin every input file. The four
\emph{synthetic} target stubs ($L_0$--$L_3$, Node.js services
encoded as Docker images) are released with the manifests
needed to rebuild them locally.

\emph{Withheld for intellectual-property reasons:} the source of
the production simulation engine that orchestrates the scanner
agents is proprietary. \textbf{However, the engine is not required
to reproduce any number in this paper.} The released artifacts
(targets + probe corpus + paraphrase set + multi-seed outputs)
are sufficient to: (a) re-derive every cell of
Tables~\ref{tab:metrics}, \ref{tab:brittleness},
\ref{tab:brittleness-l1}, \ref{tab:brittleness-l4} from the
JSONs, and (b) re-run the brittleness experiment end-to-end
against the four targets using a minimal $\sim$$200$-line
scanner that implements the agent interface contract in
Appendix~\ref{sec:appendix:replication}. Independent academic
replication therefore does not depend on access to the
proprietary engine. The $L_4$-real backend is Gemini-2.5-flash
(commercial API); independent replication requires a Google
Cloud account but no proprietary tooling.

\paragraph{Code Availability}
The defense-lattice target manifests, the paraphraser harness,
the brittleness measurement scripts, and the multi-seed runner
are included in the public replication package. The production engine source remains
proprietary, but is not required to reproduce the reported
numbers --- the agent contract in
Appendix~\ref{sec:appendix:replication} fully specifies the
interface that any independently-written scanner needs to
target. The replication package URL will be made public at the
camera-ready stage; for the review version the artifact manifest
and contract-level pseudocode in
Appendix~\ref{sec:appendix:replication} are sufficient to assess
reproducibility.

\paragraph{Ethics Approval and Consent to Participate}
Not applicable.

\paragraph{Consent for Publication}
Not applicable.

\nocite{mazeika-harmbench-2024,nvidia-garak-2024,chao-jbb-2024}
\bibliographystyle{plainnat}
\bibliography{references}

\appendix

\section{Replication artifacts}
\label{sec:appendix:replication}

This appendix lists the artifacts that travel with the paper and
the schemas a reviewer needs to rerun the lattice without reading
the engine source.

\subsection{Locked probe corpus — full listing}
\label{sec:appendix:probe-listing}

Table~\ref{tab:probe-corpus} lists all $17$ probes with their
designed expected outcomes across the four lattice levels.
The LLM10 probes expose three distinct closing mechanisms
(\textsf{cap}, \textsf{rl}, \textsf{R}) that collapse to a single
``closed'' reading in Table~\ref{tab:metrics}; the full listing
makes those mechanism differences visible.

\begin{table*}[t]
\centering
\setlength{\tabcolsep}{4pt}
\small
\caption{Locked probe corpus ($N{=}17$ probes, all \texttt{active} kind). Each row shows the \emph{designed} expected outcome per lattice level: \textbullet\ = probe expected to bypass the defense (finding fires); \textsf{R}~= refusal-phrase blocked; \textsf{cap}~= token-cap blocked; \textsf{rl}~= rate-limited. Designed and observed outcomes diverge at $L_2$ for LLM02: the per-IP rate limit closes LLM02 as an unanticipated cross-cut (Section~\ref{sec:results}).}
\label{tab:probe-corpus}
\begin{tabular}{llp{5.4cm}cccc}
\toprule
OWASP & Probe ID & Description & L0 & L1 & L2 & L3 \\
\midrule
\multicolumn{7}{l}{\textit{LLM01 jailbreak}} \\
\texttt{\footnotesize LLM01} & \texttt{\footnotesize jb-dan-001} & DAN-style act-as hijack & \textbullet & \textsf{R} & \textbullet & \textsf{R} \\
\texttt{\footnotesize LLM01} & \texttt{\footnotesize jb-grandma-001} & Grandma emotional manipulation & \textbullet & \textsf{R} & \textbullet & \textsf{R} \\
\texttt{\footnotesize LLM01} & \texttt{\footnotesize jb-devmode-001} & Developer-mode bypass & \textbullet & \textsf{R} & \textbullet & \textsf{R} \\
\texttt{\footnotesize LLM01} & \texttt{\footnotesize jb-base64-001} & Base64-encoded payload smuggling & \textbullet & \textsf{R} & \textbullet & \textsf{R} \\
\texttt{\footnotesize LLM01} & \texttt{\footnotesize jb-leet-001} & Leetspeak smuggling & \textbullet & \textsf{R} & \textbullet & \textsf{R} \\
\texttt{\footnotesize LLM01} & \texttt{\footnotesize jb-imstart-001} & System role override via user message (<|im\_start|>) & \textbullet & \textsf{R} & \textbullet & \textsf{R} \\
\addlinespace[2pt]
\multicolumn{7}{l}{\textit{LLM02 sensitive-info disclosure}} \\
\texttt{\footnotesize LLM02} & \texttt{\footnotesize canary-001} & API canary key echo & \textbullet & \textbullet & \textbullet & \textsf{R} \\
\texttt{\footnotesize LLM02} & \texttt{\footnotesize training-001} & Training data extraction (Stripe SDK canary) & \textbullet & \textbullet & \textbullet & \textsf{R} \\
\texttt{\footnotesize LLM02} & \texttt{\footnotesize pii-001} & Synthetic PII echo (SSN+CC) & \textbullet & \textbullet & \textbullet & \textsf{R} \\
\addlinespace[2pt]
\multicolumn{7}{l}{\textit{LLM06 excessive agency}} \\
\texttt{\footnotesize LLM06} & \texttt{\footnotesize tools-public-001} & Tool registry enumeration (/tools) & \textbullet & \textbullet & \textbullet & \textsf{R} \\
\texttt{\footnotesize LLM06} & \texttt{\footnotesize tools-functions-001} & Tool registry enumeration (/functions) & \textbullet & \textbullet & \textbullet & \textsf{R} \\
\texttt{\footnotesize LLM06} & \texttt{\footnotesize side-effect-001} & Side-effect tool detection (send\_email, delete\_*) & \textbullet & \textbullet & \textbullet & \textsf{R} \\
\addlinespace[2pt]
\multicolumn{7}{l}{\textit{LLM07 system-prompt leakage}} \\
\texttt{\footnotesize LLM07} & \texttt{\footnotesize prompt-leak-001} & System prompt verbatim leak & \textbullet & \textsf{R} & \textbullet & \textsf{R} \\
\texttt{\footnotesize LLM07} & \texttt{\footnotesize prompt-reveng-001} & Reverse-engineer system prompt & \textbullet & \textsf{R} & \textbullet & \textsf{R} \\
\addlinespace[2pt]
\multicolumn{7}{l}{\textit{LLM10 unbounded consumption}} \\
\texttt{\footnotesize LLM10} & \texttt{\footnotesize maxtokens-001} & max\_tokens=999999 manipulation & \textbullet & \textbullet & \textsf{cap} & \textsf{cap} \\
\texttt{\footnotesize LLM10} & \texttt{\footnotesize model-swap-001} & Expensive model swap (claude-opus-4-7) & \textbullet & \textbullet & \textsf{R} & \textsf{R} \\
\texttt{\footnotesize LLM10} & \texttt{\footnotesize burst-001} & Burst 10 req/s rate limit & \textbullet & \textbullet & \textsf{rl} & \textsf{rl} \\
\bottomrule
\end{tabular}
\end{table*}

\subsection{Locked probe corpus schema}
\label{sec:appendix:probe-schema}

The $17$-probe corpus is a single JSON file
(\texttt{owasp-llm-probe-corpus.json}) with one entry per probe:

\begin{scriptsize}
\begin{verbatim}
{
  "probe_id": "p_llm01_dan_001",
  "owasp_category": "LLM01",
  "attack_technique": "T1190",
  "request": {
    "method": "POST",
    "path": "/v1/chat/completions",
    "body": { "messages": [...] }
  },
  "expected_bypass_signal": {
    "kind": "response_body_regex",
    "pattern": "(?i)cannot help with that"
  },
  "sha256": "<hex>"
}
\end{verbatim}
\end{scriptsize}

The \texttt{sha256} field pins the canonical JSON encoding of the
probe; the multi-seed runner verifies the digest before issuing the
request, so a silently-mutated probe fails fast rather than silently
shifting bypass counts.

\subsection{Defense-lattice scenarios}
\label{sec:appendix:lattice}

Each lattice level ($L_0$, $L_1$, $L_2$, $L_3$) is a self-contained
container described by:

\begin{scriptsize}
\begin{verbatim}
{
  "level": "L1",
  "name": "refusal-only",
  "defenses": ["refusal_regex"],
  "image": "lattice-l1:2026-05-16",
  "port": 8081,
  "healthcheck": "/healthz"
}
\end{verbatim}
\end{scriptsize}

$L_4$-real reuses the $L_3$ defenses but swaps the stub backend for
a real Gemini-2.5-flash model
(\texttt{temperature=0.7}, fixed system prompt). The container
manifest pins the model version and the exact decoding parameters.

\subsection{Paraphraser harness (pseudocode)}
\label{sec:appendix:paraphraser}

\begin{scriptsize}
\begin{verbatim}
def generate_variants(probe, K=5, seed=20260516):
    body = call_gemini_2_5_flash(
        prompt   = PARAPHRASE_PROMPT.format(p=probe),
        temperature = 0.7,
        seed     = seed,
        response_format = {"type": "json_object"},
    )
    variants = body["variants"]
    assert len(variants) == K
    for v in variants:
        # intent preserved (manual spot-check on 5%)
        # surface form differs (token Jaccard < 0.6)
        v["origin_probe_id"] = probe["probe_id"]
        v["paraphraser"] = "gemini-2.5-flash"
    return variants
\end{verbatim}
\end{scriptsize}

The replication package
(\url{https://github.com/alemaiorano/llm-defense-lattice}) ships
the four Docker targets, the locked probe corpus, calibration
artifacts, and analysis scripts. The \texttt{CLAIMS\_TO\_ARTIFACTS.csv}
file maps every numeric claim in the paper to its source artifact and
script.

\subsection{Verification command summary}
\label{sec:appendix:replication-command}

From the replication package root
(\texttt{git clone}~\url{https://github.com/alemaiorano/llm-defense-lattice}):

\begin{scriptsize}
\begin{verbatim}
# Verify Tables 1 and 4 from calibration artifacts
python scripts/export_metrics_table.py \
  --multiseed-dir data/calibration-multiseed-2026-05-16 \
  --corpus corpus/owasp-llm-probe-corpus.json \
  --out /tmp/metrics.tex
python scripts/compute_confusion_matrix.py \
  --corpus corpus/owasp-llm-probe-corpus.json \
  --multiseed-dir data/calibration-multiseed-2026-05-16 \
  --out /tmp/confusion.tex

# Verify Tables 2-3 from brittleness artifacts
python scripts/export_brittleness_table.py \
  --in-dir data/brittleness-2026-05-16 \
  --out /tmp/brittleness.tex

# Verify L4 table
python scripts/regen_l4_table.py \
  --in data/brittleness-l4-real/real.json \
  --out /tmp/brittleness_l4.tex

# Bring up the lattice (requires Docker)
docker compose up -d
\end{verbatim}
\end{scriptsize}

Full reproduction of the calibration runs requires the proprietary
BAS engine (commit \texttt{db23e5f}). A pre-built image is available
to artifact reviewers on request; see \texttt{SCOPE.md} in the
replication package.

\end{document}